\documentclass[%
 reprint,twocolumn,floatfix,
 amsmath,amssymb,
 aps,
 prl,
superscriptaddress
]{revtex4-1}
\usepackage{graphicx}
\usepackage{dcolumn}
\usepackage{bm}
\usepackage{graphicx,color}
\usepackage{comment}
\usepackage{url}
\usepackage[normalem]{ulem}

\newcommand{\edit}[1]{#1}  
\begin{document}



\title{Spontaneous Domain Formation in Spherically-Confined Elastic Filaments}

\author{Tine Curk}\thanks{These authors contributed equally to the manuscript}
\affiliation{Institute of Physics, Chinese Academy of Sciences, Beijing 100190, China.}%
\affiliation{Faculty of Chemistry and Chemical Engineering, University of Maribor, 2000 Maribor, Slovenia}%
\author{James Daniel Farrell}\thanks{These authors contributed equally to the manuscript}
\affiliation{Institute of Physics, Chinese Academy of Sciences, Beijing 100190, China.}%
\author{Jure Dobnikar}%
\email{jd489@cam.ac.uk}%
\affiliation{Institute of Physics, Chinese Academy of Sciences, Beijing 100190, China.}%
\affiliation{Songshan Lake Materials Laboratory, Dongguan, Guangdong 523808, China.
}%
\affiliation{Department of Chemistry, University of Cambridge, Cambridge, CB2 1EW United Kingdom. }%

\author{Rudolf Podgornik}
\email{podgornikrudolf@ucas.ac.cn}%
\affiliation{
School of Physical Sciences and Kavli Institute for Theoretical Sciences, University of Chinese Academy of Sciences, Beijing 100190, China. 
}%
\affiliation{Institute of Physics, Chinese Academy of Sciences, Beijing 100190, China.}%

\date{\today}
\begin{abstract}
Although  the  free  energy  of  a  genome  packing  into a  virus  is  dominated  by DNA-DNA interactions, ordering  of the  DNA  inside  the  capsid  is  elasticity--driven, suggesting general solutions with DNA organized into spool-like domains. 
Using analytical calculations and computer simulations of a long elastic filament confined to a spherical container, we show that the ground state is not a single spool as assumed hitherto, but an ordering mosaic of multiple homogeneously-ordered domains.
At low densities, we observe  concentric spools, while at higher densities, other morphologies emerge, \edit{which resemble topological links}.
We discuss our results in the context of metallic wires, viral DNA, and flexible polymers. 
\end{abstract}

\pacs{Valid PACS appear here}
\maketitle

\noindent{\bf \em Introduction.} 
Spatial organization of constrained elastic filaments underlies a range of packing problems from macroscopic wires in spherical cavities~\cite{Herrmann,Bonn} to genomes in bacteriophages~\cite{Knobler2009,Smith2017}.
A theoretical interpretation of single molecule experiments on DNA packaging in viruses~\cite{Purohit2003} suggests single-domain spool-like structures with axial symmetry as a prevailing morphology.
However, recent experiments report a range of different structures with inhomogeneously-ordered genomes, defects, and phase transitions between the packings.
Cryo-electron microscopy of bacteriophages~\cite{Livolant,Leforestier-poly} at various physicochemical conditions indicates a range of possible morphologies of DNA within the capsid: from isotropically disordered, cholesteric liquid crystalline ordering with a disordered core, to uniform, concentric spools with local hexagonal order, reminiscent of the nested spools first proposed in~\cite{Hall1982}.
Measurements of DNA dynamics during packaging~\cite{Berndsen}, and intermittent ejection dynamics~\cite{Chiaruttini2010} suggest multi-domain structures. Similarly, experiments and simulations of quasi-one-dimensional elastoplastic filaments forced into spherical confinement~\cite{Herrmann,Bonn}, as well as the theoretical description of elastic filaments within the continuum polymer nematic theory~\cite{Grason2, Svensek}, indicate a multidomain structures with a mixture of local disorder and nematic order.

Coarse-grained simulations of confined semiflexible polymers mimicking DNA in viral shells have been performed at various levels of detail. DNA interactions and entropic contributions need to be considered to explain the energetics of confinement 
~\cite{kindt,Phillips2018}, but the conformation of the confined DNA 
is dictated by elasticity~\cite{petrov}.
Packing into icosahedral capsids was shown to have a strong stochastic component~\cite{Forrey}, and structures without spool-like symmetry have been reported, including multi-domain spool packing, observed when DNA is pushed into the capsid through a portal ~\cite{Rapaport}.
The effect of elastic kinks on packing has been discussed in~\cite{Myers}, while local liquid-crystalline order of compacted DNA was found to play a crucial role 
promoting the formation of knots~\cite{Marenduzzo3}. 

The slow dynamics of stiff chains in confinement implies that the configurations observed in the non-equilibrium simulations crucially depend on the applied protocol,  \textit{e.g.}, on how the external force is applied in  packaging-motor driven DNA encapsidation in bacteriophages~\cite{Phillips2018}.
It is therefore not surprising that, even in cases where the models are very similar, a wide variety of behaviours is reported. 
Here, we focus on the nature of the \textit{ground-state packing configurations}. While the present analysis thus answers a different question from previous works, its results should nevertheless provide a well-defined background scenario with which non-equilibrium configurations can be compared and assessed.


\noindent{\bf \em Theoretical Model.} We consider a spherical enclosure $\cal{C}$ with radius $R_0$ (volume $V_0$), and a hard semi-flexible cylinder with cross-section $\sigma$, length $L$, and a persistence length $l_p =  K_{\rm c}/ k_{\rm B}T$, where $K_{\rm c}$ is the bending rigidity.
The chain cannot overlap with itself nor with the confining wall and it is torsionally relaxed; its elastic energy is described by the integral $E_{el} = {\textstyle\frac{1}{2}} K_{\rm c} \int_{0}^{L}  ds/R^{2}(s)$, with $R(s)$ the local radius of curvature.

We explore the limit of a rigid, long, thin chain,
where the  elastic  energy  dominates over the  configurational entropy.
In this limit, the chain can be described by the continuum theory of polymer nematics~\cite{Svensek,Grason2}, where the polymer is described by a nematic-order vector field or local polymer ``current'' $\mathbf{t}(\mathbf{r})$~\cite{Svensek} with $\mathbf{\hat{t}}= \mathbf{t}/t$ the unit tangent vector (director) to the chain. 
The total amount of material is  $V_{p}=\int_{\cal{C}} t ~d\mathbf{r}$, with $t \equiv|\mathbf{t}|$ the local polymer density. Due to the excluded volume of the chain, $t$ is bounded by the densest (hexagonal) packing of rigid disks in two dimensions: \mbox{($t^{m} =\frac{\pi\sqrt{3}}{6}\approx 0.907$)}~\cite{inviro2016}.
The ratio between the actual amount of polymer in the capsid and its theoretical maximum, \mbox{$\nu\equiv  V_{p}/(t_m V_{0})=\frac{3\sqrt{3}\sigma^2}{8\pi R_0^3}L$}, is a key dimensionless parameter in the model. 

The polymer current is only non-zero inside the capsid
and has to be 
solenoidal ($\nabla\cdot \mathbf{t} = 0$) except at the polymer end-points (the contribution of which vanishes in the long-chain limit). 
The lowest non-trivial order in the expansion of the elastic free energy of the divergence-free polymer current is given by \cite{Svensek,Grason2}
\begin{equation}
E_{el} = \frac{2K_c}{\pi\sigma^2} {\int_V} t ~\Big(\mathbf{\hat{t}} \times (\nabla \times \mathbf{\hat{t}})\Big)^2~d{\bf r} \;,
\label{eq-Ev}
\end{equation}
where twist deformations ($\mathbf{\hat{t}} \cdot (\nabla \times \mathbf{\hat{t}})$) are by assumption excluded.
The problem is to find the 
solenoidal field $\mathbf{t}({\bf r})$ within $\cal{C}$ which minimizes $E_{el}$.

Assuming axial symmetry, the solution
is an spool~\cite{inviro2016}. 
However, relaxing the symmetry constraints might lead to more complex structures such as those seen in experiments~\cite{Leforestier-poly, Berndsen}, theory~\cite{Grason2,Svensek}, and simulations~\cite{Myers,Forrey,petrov}. 
Indeed, as we show later, a single-domain spool is \textit{almost never} the lowest elastic energy configuration.
For solutions containing more than one domain, $\mathbf{t}(\mathbf{r})$ is not continuous in space, but it can still be solenoidal if at discontinuity interfaces ${\cal S}$, 
the local field on both sides is perpendicular to the normal, $\mathbf{t} \cdot \mathbf{n}({\cal S})=0$. Multiple-domain structures should therefore be considered. 
Their advantage is flexibility in arranging the material so as to avoid highly curved regions close to the capsid center, which cannot be done in single spool structures.

\begin{figure}[t]
 \includegraphics[width=\linewidth]{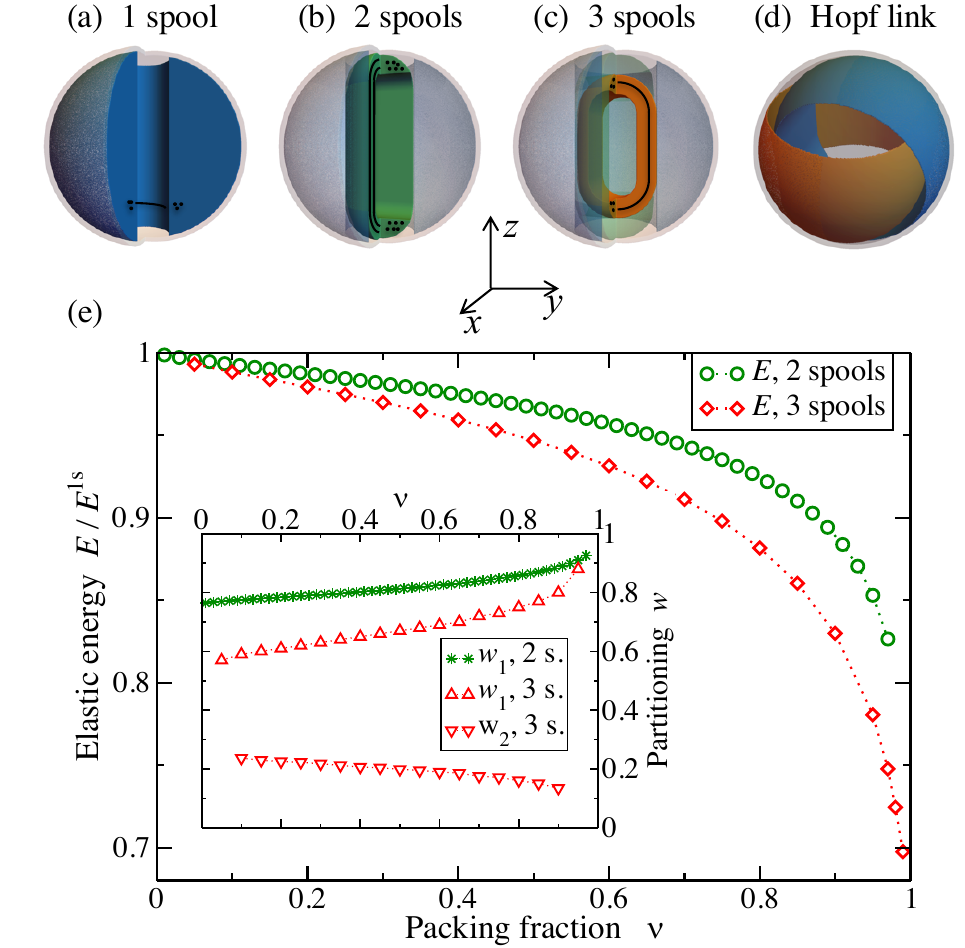}
  \caption{{\bf Theoretical model.} Visualization of the trial structures: single spool (a) with polymer-occupied volume blue, double (b) and triple (c) spools (only the inner spools shown in green and orange, respectively), and the Hopf link (d). The structures shown are the solutions with optimal partitioning $w$ at $\nu=0.9$. 
  (e): The optimal ratio of elastic energies $E_{el}^{2s, 3s}/E_{el}^{1s}$ as a function of $\nu$ for double spools (green circles) and triple spools (red diamonds).
  {\em Inset:} The optimal partitioning as a function of $\nu$ for double spools ($w_{1}$, green stars) and triple spools ($w_{1,2}$, red triangles). 
  }
  \label{fig-theory}
\end{figure}

With this in mind, we assume that the field  is partitioned into $n_d$ domains, $\nu=\sum_{1}^{n_d} {\nu_i}$, with relative sizes $w_i\equiv \nu_i/\nu$.
For $n_d\le 3$, we construct candidate structures and search for the optimum partitioning $\{m_i\}$ at several packing fractions $0<\nu<1$.
Inspired by observations~\cite{Herrmann, Bonn,svensek-EPL,Rapaport, Forrey, petrov}, we consider four classes of trial structures depicted in Fig.~\ref{fig-theory}: single spools (a), double spools (b) comprising two nested tori with perpendicular axes, triply nested spools (c) with perpendicular axes, and---deviating from the nested-spool paradigm---Hopf links (d),  with a symmetric partitioning into two equivalent, perpendicular, topologically-linked tori.
We assume that the field density is $t_m$ within the defined domains and zero elsewhere and approximated \eqref{eq-Ev} by a sum of integrals over independent domains. We neglect the short segments connecting the domains, arguing that this contribution is negligible in the thin, long chain limit. 

For a single spool, 
$\mathbf{\hat{t}}^{1s} = t_m \left({-y}, {x}, 0\right)/\sqrt{x^2+y^2}$, we have~\cite{Purohit2003}:
\begin{equation}
E_{el}^{1s} (\nu, R_0) =  \frac{\pi^2 K R_0}{\sqrt{3}}   \Biggl[ - {\nu}^{1/3} + \textstyle{\frac{1}{2}}\ln  \frac{1+\nu^{1/3} }{1-\nu^{1/3} }\Biggr ]\;.
\label{eq-Et1s}
\end{equation}
The energy diverges for $\nu\to 1$, due to diverging curvature in the center of the capsid. 
The Hopf link comprises two perpendicular spools with a partitioning dictated by symmetry ($w_{1}=0.5$) and
$E_{el}^{\rm hl} \left(\nu, R_0\right) \approx 2 E_{el}^{1\rm s} \left(\nu/2, R_0\right)$.
In the double spool, the outer domain ($\nu_1$) is the same as above, and the inner spool ($\nu_2$) with radius  $R_1=R_0\sqrt{1-\nu_1^{2/3}}$ is composed of two small hemi-toroidal caps of combined length proportional to $\nu_2^{\rm t}$, connected by straight sections ($\nu_2^{\rm s}$): $\nu _2=\nu_2^{\rm t} + \nu_2^{\rm s}$.
The elastic energy is localized in the hemispheres and is determined by $\nu_2^{\rm t}$ alone (see section SI(B)):
\begin{equation}
E_{el}^{2\rm s} (\nu_1, \nu_2, R_0) =  E_{el}^{1\rm s} (\nu_1, R_0) + E_{el}^{1\rm s} (\nu_{2}^{\rm t}, R_1).
\label{eq-Et2s}
\end{equation}
Interestingly, at a given $\nu$, the ratio \mbox{$E_{el}^{2\rm s}/E_{el}^{1\rm s}$} is completely determined by the partitioning of the material $w_1$. 
We calculate the optimal $w_1$ for several values of $\nu$ by minimizing \eqref{eq-Et2s}. 
The calculation for triple spools is analogous: $\nu=\nu_1+\nu_2+\nu_3$, and the elastic energy a sum over two outer spools \eqref{eq-Et2s} and an innermost spool with an elaborate shape (see section SI(C)): $E_{el}^{3\rm s} \left(\nu_1, \nu_2,\nu_3, R_0\right) =  E_{el}^{2\rm s} \left(\nu_1, \nu_2, R_0\right) + E_{el}^{3^{\rm rd}\rm spool} \left(\nu_1,\nu_2,\nu_3, R_0\right) $.
The optimal structure is obtained by minimizing over $w_1$ and $w_2$. 

Fig.~\ref{fig-theory}e) shows the ratio of the elastic energies as a function of $\nu$ for optimal double- and triple-spool configurations. We note that both are always lower in energy than a single spool, and that the triple spool wins over the double spool.
\edit{Adding a fourth spool in an analogous manner does not reduce the energy further (see online SI for details).} 
 
Our best Hopf link structure (Fig.~S4) can have lower elastic energy than the single spool only at extremely low $\nu$ and even then by a tiny margin, but never wins over double or triple spools.

The robust conclusion within our analytical model is that in the long, thin, and hard cylinder limit, multiple domains are favored over a single-domain.
In principle, this may differ if we consider spools with non-orthogonal axes, or any other, better arrangements not included in our theoretical considerations.
Moving away from this limit (\emph{i.e.}, finite $\sigma$, repulsive interactions other than excluded volume), the volume accessible to the polymer will be modified, which can result in a domain-wall energy penalty, disfavoring multi-domain arrangements.
At finite temperature, entropic terms due to conformational fluctuations of the chain also become important.
We will address all of these issues with computer simulations.

\noindent{\bf \em Simulation Model.} We consider a long elastic filament of diameter $\sigma$ confined to a sphere with radius $R_{0}=5\sigma$.
We model the elastic filament as a sequence of $N$ soft, repulsive beads interacting with each other and the confining wall \textit{via} the \mbox{Weeks--Chandler--Anderson} (WCA) repulsion~\cite{weeks1971role}.
We use standard LJ units, expressing energies in $\epsilon$ and lengths in $\sigma$.
The beads are bonded by a two-body stretching term ${V_{b} = 16\,\Gamma \left(r-1\right)^{2}}$ and a three-body bending term ${V_{\theta}= \Gamma \left(1+\cos{\theta}\right)}$, where $r$ is the distance between two consecutive beads, and $\theta$ the angle between consecutive bond vectors.
The ratio 16 follows the exact result for elastic rods \cite{Rudi_elastic_rods}, and the prefactor $\Gamma$ determines the chain rigidity, $\Gamma=T^{*}l^{*}_p$, with $l^{*}_p$ the dimensionless persistence length and $T^{*}\equiv k_B T/\epsilon$ the reduced simulation temperature.
Changing $T^*$ or $l^*_p$ has the same effect on the elastic energy; therefore, by fixing $\Gamma$, we can simulate polymers with different rigidity by changing the simulation temperature~\footnote{This is not entirely true, since the LJ term does not scale with $\Gamma$.
However, in the rigid limit the effect of repulsive interactions on the persistence length is negligible.}.
In our simulations, we fix $\Gamma=1000$, so at low temperatures we are modeling fairly rigid chains (at $T^*=1$, $l_p/\sigma=1000$) and at $T^*\approx 40$ ($l_p/\sigma\approx 25$) the parameters correspond to DNA.

We performed $NVT$ replica-exchange molecular dynamics simulations~\cite{ParallelT_1997,sugita1999replica} of chains of up to $N=250$ beads using the \texttt{LAMMPS} package~\cite{plimpton1995fast}. 
We simulated 128 replicas with geometrically-distributed temperatures $T^{*}(n) = 1.045^n$ regulated by a Nos\'e--Hoover thermostat for up to $2\times10^{9}$ time steps, attempting exchanges between replicas every $10^{3}$ steps, and dumping snapshots every ${10^{6}}$ steps. 
We employ a variable time step which is dynamically determined such that no bead moves by more than $0.005 \sigma$ per time step~\footnote{The time step is thus smaller than in standard LJ simulations because we need to resolve the time scale of bond vibrations.}.
The polymer packing fraction is $\nu=3N(R_0-0.5)^{-3}/16t_m$, where the accessible capsid volume is rescaled due to finite chain thickness. 
Thus, for $100\le N\le 250$, $0.23 \le \nu \le 0.57$. 
For larger packing fractions (we tried up to $N=350$, $\nu=0.8$), slow kinetics prevents equilibration. 

\label{sec:locmin}
We \textit{a posteriori} minimized the energy of each snapshot using the limited-memory Broyden--Fletcher--Goldfarb--Shanno (l-BFGS) algorithm ~\cite{nocedal1980updating, liu1989limited} as implemented in the \texttt{pele} package~\cite{stevenson2012pele}, converging the root-mean-squared gradient to a tolerance of $1\times10^{-3}\frac{\epsilon}{\sigma}$. We further optimized the energy of selected structures with a combination of random displacement, reptation, and l-BFGS minimization cycles~(see algorithm S1).

\begin{figure}[t!]
\includegraphics[width=\linewidth]{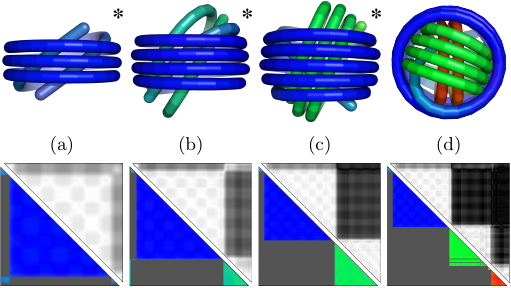}
\caption{{\bf Spool morphologies} at $N=100$ (a), 150 (b), 200 (c), and 250 (d).
Structures are drawn as tubes connecting bonded monomers, and coloured according the to direction of $\mathbf{\hat{b}}$.
\edit{Putative ground-state structures are denoted by asterisks.}
Bottom: binormal angle correlation (upper-right) combined with domain decomposition (lower-left).
$\left(m_{1},m_{2},m_{3}\right)=\left(84,13,3\right)$--$\left(107,35,8\right)$--$\left(114,73,10\right)$--$\left(115,77,34\right)$ from (a-d).
}
  \label{fig:sam_a.pdf}
\end{figure}

Our theoretical analysis suggests that the elastic energy of a polymer packed into a sphere can be minimized by segregating into up to three distinct spools.
Chain segments that belong to the same spool lie in almost parallel planes. As such, we can characterize the degree to which two beads belong to the same spool domain by comparing the binormal vectors at their positions. The binormal to the chain is given by $
\mathbf{\hat{b}}=\mathbf{\hat{t}}\times\mathbf{\hat{n}}
=\frac{\mathbf{r^{'}}\left(s\right)\times\mathbf{r^{''}}\left(s\right)}{\left|\mathbf{r^{'}}\left(s\right)\times\mathbf{r^{''}}\left(s\right)\right|}
$,
where $\mathbf{r}$ is the position of the particle, $s$ is the arc length along the chain, $\mathbf{\hat{t}}$ is the unit tangent vector, and $\mathbf{\hat{n}}$ is the unit normal vector. A simple way to visualize the structure of a compacted chain is to plot a correlation plot of the angles between binormal vectors at each pair of particle centers (e.g.~Fig.~\ref{fig:sam_a.pdf}). We identify the principal domains by searching for cliques using the \texttt{networkx} package~\cite{hagberg2008exploring}. 

The size of a domain $D_i$ is denoted as $m_{i}=w_i N$. The largest domain, $D_{1}$, is the largest set of beads such that the angle between the binormal to the chain for every pair of beads is less than $30^{\circ}$ (see Fig.~S5).
The $n^{\mathrm{th}}$ largest domain, $D_{n}$ is then the largest such set disjoint with the union of all larger domains, $\bigcup_{i=1}^{n-1} D_{i}$. 

\begin{figure}[t!]
\includegraphics[width=\linewidth]{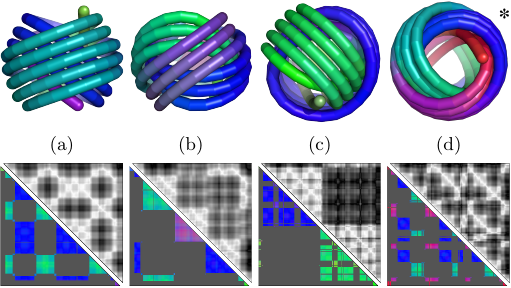}
\caption{{\bf Link motifs.} $N=200$: \edit{Hopf link (a) and link L6n1 (b); $N=250$: Hopf link (c) and link L6n1 (d)}. \edit{(d) is the putative ground state for $N=250$}.
Bottom row: binormal correlation/domain decomposition plots.
From (a-d):
\edit{$\left(m_{1},m_{2},m_{3}\right)=
\left(91,73,16\right),\left(61,60,50\right),\left(89, 80, 29\right),\left(83, 60, 55\right)$.}
}
  \label{fig:sam_b.pdf}
\end{figure}

\noindent{\bf \em Results.} Analysis of the partitioning at different conditions reveals the following picture. In the lowest-temperature replicas we observe spool morphologies for all chain lengths (see Fig.~\ref{fig:sam_a.pdf}). 
The ground state of a short chain ($N=100$) is a single spool whose ends curve inward toward the sphere center (Fig.~\ref{fig:sam_a.pdf}a). 
Simulations of 150- and 200-bead chains show the emergence of double-spool structures (Fig.~\ref{fig:sam_a.pdf}b-c), while the variety of low-energy morphologies sharply increases at $N=200$. 
This trend is illustrated by histograms of $m_{1}$ (see Fig.~S6).
%
\edit{A double spool is the putative ground state structure for 200 beads, but other morphologies, resembling topological links, are also observed (Fig.~\ref{fig:sam_b.pdf}a-b)~\footnote{We stress that these structures only \textit{resemble} topological links, in a superficial sense; the simulated chains are not closed, nor, in general, are the chain segments that constitute the `components' of the links contiguous.}.}
The spool-like structures are consistent with our theoretical assumptions (Fig.\ref{fig-theory}a-c).
We also observe a hierarchical mass distribution $m_1 \gg m_2 \gg m_3$ characteristic of the optimal analytic solutions, distinguishing spools from other morphologies, though quantitative comparison of $w_{i}$ between theory and simulations is difficult due to the sensitivity of the partitioning to the model parameters.

Upon increasing the polymer length to 250 beads, a triple spool (Fig.~\ref{fig:sam_a.pdf}d) becomes more favourable than double spools, but non-spool-like morphologies are more commonly observed. 
\edit{The structure with the lowest \emph{elastic} energy ($V_{b}+V_{\theta}$) is a Hopf link-like \mbox{morphology (Fig.~\ref{fig:sam_b.pdf}c)}, some $36\epsilon$ lower in total energy than the triple spool. The putative ground state, another $\approx59\epsilon$ lower in total energy, resembles three linked rings (L6n1 in Dowker-Thistlethwaite notation~\cite{doll1991tabulation}; Figs.~\ref{fig:sam_b.pdf}d, S7).
For the dependence of the lowest energies of these motifs on chain length, see Fig.~S8.}
Twisted packings have been proposed as models of toroidal aggregates of long chiral molecules such as DNA~\cite{kulic2004twist,Charvolin2008}; even in the absence of intrinsic twist, such packings were shown in~\cite{Grason2} to be preferred at high densities, as they can fill the sphere without disclinations or voids.
Space-filling considerations may also drive the transition to twisted, link motifs we see here.

\begin{figure}[b!]
\includegraphics[width=\linewidth]{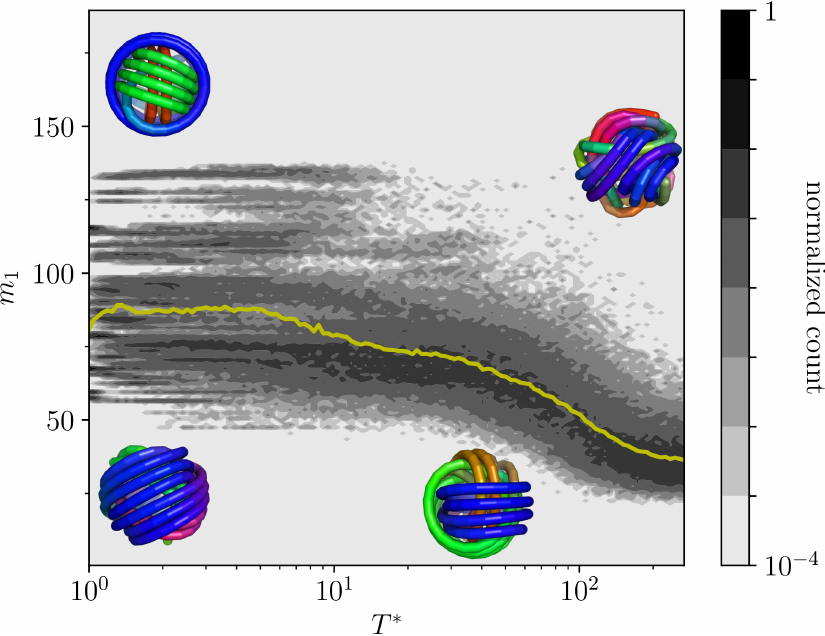}
\caption{{\bf Temperature dependence.}
The probability distribution of largest cluster size $m_{1}$ as a function of $T^{*}$ for simulations with chain length $N=250$. \edit{The persistence length is $l_p^*=1000/ T^{*}$.} 
The yellow curve is the mean value $\langle m_{1}\rangle (T^*)$. 
The snapshots are representative configurations at the stated temperatures: triple spool and \edit{L6n1 link ($T^{*}=1$)}, partially-ordered Hopf link ($T^{*}\approx40$), and disordered random coil ($T^{*}\approx270$). 
} \label{fig:m1_temperature_250.pdf}
\end{figure}

The degree of binormal ordering decreases with temperature (Fig.~\ref{fig:m1_temperature_250.pdf}).
The distribution of largest domain sizes $m_1$ for a chain of 250 beads is visualized as a heat map for temperatures $1\le T^*\le 270$.
spool structures ($m_{1}\gtrsim 100$) are rarely observed beyond $T^{*}=10$.  
For $10 < T^{*} < 40$, there is less order, but there are clear indications of \edit{link motifs}.
At high temperatures ($T^{*}>100$), while patches of order are retained near the boundary wall---due to local hexagonal packing---large ordered domains seldom emerge. 

\noindent{\bf \em Discussion \& Conclusions.} 
According to our results, at low temperature, spherical confinement is sufficient to order the majority of the filament into tight spools. 
We predict a variety of multi-domain morphologies such as nested spools and \edit{links of two or three components} competing in the low temperature regime.
Their energetic ordering may be sensitive to the interplay between elastic and repulsive forces and other microscopic details of the system.
Our low-temperature results can be compared with previous experiments and simulations of packing metallic wires into spheres~\cite{Herrmann,Bonn}, where ring-like coiling was observed, reminiscent of our nested spools.
However, those systems are in the athermal limit, and the structures adopted by wires that are pushed into the system with large forces are protocol-dependent.
In our work, we avoid this protocol dependence by employing the replica exchange method; as such, our low-temperature structures may describe kinetically-inaccessible ground states of their systems.
Furthermore, in the case of metallic wires, the confinement size was varied, with a focus on relatively large containers.
In our work, we fixed the enclosure size to $R_0=5\sigma$, which is at the very lowest end of their parameter range, where little data is available about the observed structures.
Further work is needed to explore the packing as a function of the capsid size.

On the other hand, the chosen capsid size is appropriate for modeling DNA packing into viral capsids.
For DNA, $\sigma \approx 2$ nm, which means that the viral capsid corresponding to our model would be about 20 nm in diameter, \edit{which is at the small end of the range for viral capsids in nature (20 - 200 nm).}
Given a persistence length of 50 nm, the DNA regime is described by simulation temperatures $T^{*} \approx 40$.
The structures we observe in this regime are only partially ordered, retaining only some features of low-temperature motifs.
Approaching the high-temperature limit (the flexible-chain regime), we eventually observe little structure, in agreement with previous studies~\cite{Cacciuto2006}.

Projecting our model on real systems, we acknowledge that system-specific details may influence packing motifs.
Specific DNA--capsid interactions enhance the ordering of DNA in viral capsids \textit{via} built-in folding pathways~\cite{Twarock:2018}. Structures incompatible with the $I_{h}$ symmetry of many viral capsids \edit{(\textit{e.g.}, the orthogonal Hopf link with $D_{2d}$ symmetry)} might be disfavoured.
Moreover, the kinetic pathway of motor-driven insertion likely selects spool-like morphologies~\cite{Rapaport} even if they are metastable.
Similarly, capsid ejection dynamics might depend crucially on the topology of the DNA structure, with knotted~\cite{Marenduzzo201306601} or linked morphologies possibly obstructing the ejection.

\noindent{\bf \em Acknowledgements.} 
We acknowledge enlightening discussions with Arman Boromand, Daan Frenkel, Erika Eiser, Erik Luijten, and William Gelbart. The work was supported by the EU's Horizon 2020 Program through grants ETN 674979-NANOTRANS and FET-OPEN 766972-NANOPHLOW, \edit{by the ARRS grant Z1-9170}, by the 1000-Talents Program of the Chinese Foreign Experts Bureau, and by the Chinese National Science Foundation through grants 11874398, 11850410443, and 21850410459.

\bibliography{refs_polypack} 

\end{document}